\documentclass[prl,twocolumn,superscriptaddress,footnote]{revtex4-1}
\usepackage{graphicx}
\usepackage{float}
\usepackage{epsf,latexsym,bbm,euscript}
\usepackage{amssymb,amsmath}
\usepackage{mathtools} %used in this file for the command \coloneqq -> :=
\usepackage{xcolor}
\usepackage{soul}
\usepackage{ulem}
\def\bra#1{\langle{#1}|}
\def\ket#1{|{#1}\rangle}
\def\braket#1{\langle{#1}\rangle}

{\catcode`\|=\active
  \gdef\Braket#1{\begingroup
\mathcode`\|32768\let|\BraVert\left<{#1}\right>\endgroup}}
\def\BraVert{\egroup\,\mid\,\bgroup}

\newcommand{\defeq}{\vcentcolon=}
\newcommand{\eqdef}{=\vcentcolon}

\def\6{\langle}
\def\9{\rangle}

\newcommand\vx{{\vec{x}}}
\newcommand\va{{\vec{a}}}
\newcommand\vom{{\vec{\omega}}}

\newcommand\vsi{{\vec{\sigma}}}

\def\half{\tfrac{1}{2}}

\newcommand{\pad}{\partial}
\newcommand{\be}{\begin{equation}}
\newcommand{\ee}{\end{equation}}
\newcommand{\ba}{\begin{eqnarray}}
\newcommand{\ea}{\end{eqnarray}}

\newcommand{\tcblack}{\textcolor{black}}

\begin{document}

\title{Weak-value magnetometry for precision tests of fundamental physics}
%\date{Today}
\author{Sounok Ghosh}
\affiliation{Department of Physics, Indian Institute of Technology-Bombay, Powai, Mumbai 400076, India.}
\author{Leong-Chuan Kwek}
\affiliation{Centre for Quantum Technologies, National University of Singapore, 3 Science Drive 2, 117543 Singapore, Singapore.}
\affiliation{Institute of Advanced Studies, Nanyang Technological University, Singapore 639673.}
\author{Daniel R. Terno}
\affiliation{Department of Physics \& Astronomy, Macquarie University, Sydney NSW 2109, Australia.}
\affiliation{Shenzhen Institute for Quantum Science and Engineering, Southern University of Science and Technology, Shenzhen 518055,  China.}
\author{Sai Vinjanampathy}
\email{sai@phy.iitb.ac.in}
\affiliation{Department of Physics, Indian Institute of Technology-Bombay, Powai, Mumbai 400076, India.}
\affiliation{Centre for Quantum Technologies, National University of Singapore, 3 Science Drive 2, 117543 Singapore, Singapore.}

\begin{abstract}

Progress in testing fundamental physics relies on our ability to measure exceedingly small physical quantities. Using  a $^{40}$Ca$^{+}$ trapped ion system as an example we show that an exceedingly weak synthetic magnetic field (at the scale of $10^{-19}$ T) can be measured with current technology. This improved sensitivity can be used to test the effects of spin coupling that affect the equivalence principle and, if present, may impact the performance of the proposed entangled optical clocks arrays.
\end{abstract}

\maketitle

\emph{Introduction.---} Advances in our ability to manipulate and control light and matter  {interactions}  enable{d experimental  demonstrations of the counterintuitive properties of quantum mechanics.
Today they form the basis of the emergent quantum technologies and the concomitant improvements in metrology   {facilitate  novel} tests of the fundamental physics \cite{guide:bt,aom:rmp,opt-mag}.
For example,  some of the most sensitive methods of measuring magnetic
fields are based on interactions of light with atomic vapor \cite{opt-mag}. These optical magnetometers are used for practical measurements of magnetic fields and also for tests
of monopole-dipole couplings,  searches for dark matter and Lorentz-violating interactions \cite{aom:rmp,opt-mag}.

Weak values, originally introduced as a ``new kind of value for a quantum variable'' \cite{aav,Kedem,torres2016weak,kedem2014obtaining} have recently advanced from the discussions of
 quantum foundations to practical metrology \cite{weak:pr,weak:rmp,Genovese2016}.
A large weak value essentially amplifies a signal  and allows a sensitive estimation of small evolution parameters.
This weak value amplification (WVA) comes  {at a} cost, namely a decreased success probability  that may  {erase any gains} arising from the amplification \cite{weak:rmp}.
{Nevertheless,  a judicious use of  {the} advantages  {of} this method can offer robustness against
  various types of noise, (e.g., thermal dissipation, damping and $1/f$ noise), allowing for a significant  improvement in {the sensitivity of the measured signal} with ``relatively modest'' \cite{weak:rmp} experimental resources.

 {Using} a $^{40}$Ca$^{+}$ ion trapped in a linear Paul trap with the internal (electronic) spin degree of freedom as a sensor, we  {present}
 two systematic ways to leverage the advantages of the WVA for its use in metrology.
 The small effect of the internal spin coupling to the weak magnetic field (actual or effective)
  is amplified and stored in the vibrational mode of the trapped ion and can be read out with ease.
   {First}, by  {using} dynamical decoupling schemes to combat decoherence,
   while preserving the amplification, the effect of noise  on the spin can be sufficiently mitigated. Second, we employ  a quantum flywheel, that was originally proposed as a device for extracting work from a  quantum heat engine \cite{ldk:16}
   to accumulate the generated signal.  Even with a non-optimal proof-of-the principle implementation of these procedures, our simulations  show improvements  that allow us to perform enhanced tests of fundamental physics with existing technology.
We begin by introducing spin-gravity coupling as a concrete example of such a test and
comment on the strength  {and potential effects of various terms}. Then  {we introduce}  our WVA protocol and discuss  {the simulation of a realistic experiment to detect an exceedingly weak synthetic magnetic field} and present our conclusions.

\emph{Spin-Gravity Coupling.---} Laws of gravity and especially the equivalence principle(s) are among the oldest targets of continuously improving precision tests \cite{aom:rmp,opt-mag,will-lrr,ni:10,will:93,kr:11}.
The Einstein equivalence principle (EEP) is the foundation  of general relativity (GR) and all other metric theories of gravity \cite{will-lrr,ni:10,will:93}.
We leave aside the question of if and how the EEP is violated/modified by quantum mechanics  in the absence of exotic interactions  \cite{casc:83,lamer:93,d:04,zb:18}.

Instead we take a pragmatic approach of the effective field theory.  \tcblack{The action $S$ for} the fermionic sector of the gravitationally
coupled  the {standard} model extension (SME) \cite{kt:11,t:16}, $S=S_G+S_{LV}+S_\psi$, is comprised of three terms. The term $S_G$ is the standard  Einstein-Hilbert action, possibly
supplemented by additional curvature-dependent
 terms such as those of $f(R)$ theories. The term $S_{LV}$ governs the dynamics of the coefficient  fields causing Lorentz
violation. The fermionic term is derived from the Lagrangian density
\be
{\mathcal{L}_\psi}=\half ie_{a}^\mu\bar\psi\Gamma^a\overleftrightarrow{D}_\mu\psi -\bar\psi M\psi.
\ee
Here $e_{a}^\mu$ are the tetrad vectors %\footnote{The Greek indices refer to the local coordinate $x^\mu$, and the Latin indices label the components with respect to a local Lorentz tetrad $e_a$},
  the covariant derivative acts on the fermion field as
\be
D_\mu\psi\defeq \pad_\mu\psi+\tfrac{1}{4}i\omega_\mu^{ab}\sigma_{ab}\psi,
\ee
the spin-connection $\omega_\mu^{ab}$ may depend on torsion \cite{ham:rpp}, $\sigma_{ab}\defeq \tfrac{i}{2}[\gamma_a,\gamma_b]$, and the standard Dirac matrices  $\gamma^a$ and the mass $m$
are the first terms in the expression $\Gamma^a$ and $M$, respectively. Both $\Gamma^a$ and $M$ also include a variety of possible exotic terms.

\emph{The Hamiltonian.---}
   The benchmark for identification of the new physical effects is provided by a non-relativistic limit of the Dirac equation on curved background.% It also takes into account the effects of the Earth rotation.
 The action $S_\psi$ where that involves only the standard GR coupling results in \cite{bd:82,hn:90}
 \be
 (i\hbar \gamma^\mu D_\mu -mc)\psi=0, \label{Dirac-st}
 \ee
where  $\gamma^\mu\defeq e^\mu_a\gamma^a$, the connection in  $D_\mu$ is metric-compatible and we explicitly track $\hbar$, $G$ and $c$.
If one considers energy levels of a bound non-relativistic system on Earth the new physics should appear as small effects that cannot be extracted from  Eq.~\eqref{Dirac-st}.

  The leading terms of the resulting Hamiltonian of a free spin-$\frac{1}{2}$ particle
that take into account the effects of  rotation of the reference frame with  angular velocity $\vec\omega$ and   acceleration $\vec a$ (or a uniform gravitational field)
can be represented as
\be
H=H_\mathrm{cl}+H_\mathrm{rel}+H_\sigma+H_\mathrm{ext}.
\ee
 The first three terms on the right hand side are obtained by performing the    standard Foldy-Wouthuysen transformation and taking the non-relativistic limit \cite{hn:90}.
  The term
\be
 H_\mathrm{cl}=\frac{\vec p\,^2}{2m}+ m\va\cdot \vx -\vec{\omega}\cdot L,
 \ee
 where we removed the rest mass, represents the  Hamiltonian of a free non-relativistic particle in a non-inertial frame.
 The term  $H_\mathrm{rel}$ describes the higher-order relativistic corrections
 that do not involve spin. The leading spin-dependent terms
  \be
  H_\sigma=-\half\hbar\vom\cdot \vsi  +\frac{\hbar}{4mc^2}\vec\sigma\cdot(\va\times \vec p), \label{hn12}
   \ee
where $\vsi$ are the Pauli matrices, are due to Mashhoon \cite{m:88} and Hehl and Ni \cite{hn:90}.

 Finally, the term
\be
H_\mathrm{ext}=\frac{\hbar k}{2c}\va\cdot\vsi \label{PO}
\ee
represents the unconventional spin-acceleration (or spin-gravity) coupling.  It is a limiting form of the simplest phenomenological addition to the Dirac equation that breaks the equivalence principle and parity invariance
\cite{mp:62,lo:64,p:78}.
We refer to it as the Peres term.
For the value $k=1$       it results from the Eriksen and Kolsrund version \cite{ek:60} of the  Foldy-Wouthuysen  transformation \cite{o:01}.
 Since the resulting Hamiltonian is not invariant under the standard non-relativistic parity \tcblack{transformation} this term is  likely  a mathematical artefact.
  However,   Eq.~\eqref{PO}
 arises in the non-relativistic limit of various gravitational SMEs. For example, in the Moody-Wilczek-Dobrescu-Mocioiu formalism \cite{aom:rmp,mw:84,dm:06} this is the limiting form of the
  monopole-dipole potential $\mathcal{V}_{9,10}(r)$
 that is generated by a light pseudoscalar filed with the effective range exceeding the radius of the Earth.

  \emph{Strength Estimates.---}
 The spin-dependent terms are small under normal conditions. On the Earth surface $\hbar g/c=2.15\times 10^{-23}$ eV, which is equivalent to the effective magnetic field of $3.7\times 10^{-19}$ T.
 This is still several orders of magnitude below the peak sensitivity of the optical magnetometery \cite{opt-mag}.
% The spin-rotation effect was demonstrated  by using neutron polarimetry in rotating magnetic field \cite{dsh:15}, and there are indications  of importance of the  Mashhoon term in  several systems \cite{pl:02,lp:13}.
The Mashhoon term is significantly larger than the Peres term with $k=1$, since $\omega c/g=2.22\times 10^3$. The Mashhoon term is about an order of magnitude stronger onboard of the
satellites that are planned, e.g., to carry entangled optical clocks aiming to establish the next level of precision and stability of $10^{-18}-10^{-20}$ \cite{komar:14}. In searches for the direct spin-gravity coupling
 of Eq.~\eqref{PO}   \tcblack{\tcblack{effect of rotation}} is approximately cancelled by having the spin to precess
about an axis nearly parallel (or antiparallel) to that
of the Earth's rotation \cite{venema:92},  and the residual phase is removed during the data post-processing \cite{venema:92,gemmel-der:10}.

This synthetic Zeeman effect will be manifest as a small perturbation on the optical clock levels.     The working transition frequencies correspond to $0.5-2.5$ eV energy gap and
are established with a fractional uncertainty $\delta\nu/\nu$ that is within
the range $0.6-250 \times 10^{-15}$   \cite{oclock:10}. Rotation   introduces additional shifts to the energy levels with non-zero total spin and/or orbital angular momentum of the order of $10^{-15}$ eV (and one order of magnitude larger for    a satellite with  an orbiting period of two hours).
The inertial effects are  much more serious      problem for the standard atomi clocks that operate on the hyperfine transition. For example,
 the basis for the standard clock hyperfine splitting (in Cs, between $F=4$ and $F=3$ hyperfine sublevels of ${}^2S_{1\!/\!2}$ is 9,192,631,770Hz or
 approximately $3.8018\times 10^{-5}$ eV).

 Depending on the particle tested and the experimental method the limits on $k$ in Eq.~\eqref{PO}
 range between $10-10^4$ \cite{venema:92,bound:17,lim1,lim2}.  The use of WVA can improve these bounds.

%Since the Hehl-Ni term (the second term in Eq.~\eqref{hn12}) is suppressed by the additional factor of $v/c$, this discussion leads us to consider the following two terms
%\be
%H_\mathrm{spin}= -\half\hbar\vom\cdot \vsi + \frac{\hbar k}{2c}\va\cdot\vsi,   \label{hspin}
%\ee
%as the spin-dependent interaction in our study. \tcr{ We focus in this manuscript on the measurement of the Peres term, since it is smaller than the Mashhoon term and hence more difficult to detect. Since all three terms appear as synthetic magnetic fields with varying strengths, our technique can be modified readily to measure the stronger Mashhoon term. The conventional term, being weaker than the Peres term, requires more precise measurements.}

\emph{Weak Value Amplification.---}
The WVA technique \tcblack{derives} from the standard von Neumann measurement procedure {similarly to the construction of the positive operator-valued measure, but specifically adapted to small values of the measured quantities}
\cite{aav,duck,weak:pr,weak:rmp,knee2016weak}. The  technique proceeds
by coupling the quantum system with the measurement apparatus via a generic interaction Hamiltonian of the form $H=\varepsilon A\otimes M$, where $\varepsilon$ is the coupling, $A$ is the
system's observable to be measured and \tcblack{the operator} $M$ is an operator describing the other subsystem, often called the ``meter".
\tcblack{The} quantum system  S  and the meter M \tcblack{are} initially in a \tcblack{product} state \tcblack{$\ket {\Psi_{i}} = \ket {s_{i}}\otimes \ket {m_{i}}$}.

 The combined system evolves to
	\be
	|\Psi_f\9= e^{-i\gamma A\otimes M}  |\Psi_i\9\eqdef U(t)|\Psi_i\9,
	\ee 
where $\gamma=\varepsilon t$ is assumed to be small.

{The evolution is}  followed by a post\tcblack{selection to the state} $\ket{s_f}$ \tcblack{of the system}, transforming $\ket{\Psi_{i}}$ to
%\begin{equation}
%    \ket{\phi}_{f}= \braket{\psi_{f}|U(t_{f};t_{0})|\Psi_{i}} \approx  %\braket{\psi_{f}|\mathbb{I}_{SM}+-i\gamma A\otimes q|\Psi_{i}}
%\end{equation}
\begin{equation}
\ket{m_f}=  \braket{s_{f}|s_{i}}\ket{m_{i}}-i\gamma M\braket{s_{f}| A|s_{i}}\ket{m_{i}}\propto   e^{-i\gamma A_{w}M}\ket{m_{i}},
\end{equation} 
where $\mathcal{A}_{w}=\braket{s_{f}| A|s_{i}}/\braket{s_{f}|s_{i}}$ {is called the weak value}. {This occurs with the probability  {$p_{f}:=\vert\braket{s_{f}|s_{i}}\vert^2$}}.
{For nearly orthogonal states $\ket{s_{f}}$ and $\ket{s_{i}}$, we
see an amplification in the $M$-generated translation in $\ket{m_i}$, registered in the so-called ``kicked" state $\ket{m_f}$. Several experiments report WVA in the laboratory settings for different physical system \cite{hosten2008observation,li2018phase,Wang,Goswami:14}.

Though \tcblack{WVA} is known to outperform conventional measurements \cite{Magana-Loaiza2017} in some cases, it
 offers an advantage only if the information discarded in the post-selection is negligible. Here information is quantified by the Fisher information (see Appendix C for details) wherein a comparison is made between the information available in (a) the initial system-meter state, (b) the total state after a successful postselection event $\ket{s_f}\otimes\ket{m_f}$ and (c) the post-selected meter state $\ket{m_f}$. We denote the total quantum Fisher information in the initial state as $F_T(g)$. The quantum Fisher information following post-selection $F_{ps}(g)$ can be written as $F_{ps}(g) = F_{m}(g) + F_{p_f}(g)$, where $F_{m}(g)$ is the quantum Fisher information available in the meter state and  {$F_{p_f}(g)$} is the classical Fisher information from the post-selection probability distribution \cite{Zhang,Combes,Alves,Knee}.

The weak coupling regime for WVA is defined as $g|A_{w}|\Delta \ll 1 $ where $\Delta$ is the standard deviation of the distribution of the initial eigenvalues of $M$ \cite{weak:pr,weak:rmp}. 
In this regime, though there is a loss of statistics, the discarded data contains less and less information, i.e.,  {$F_{p_f}/F_{ps} \ll 1$}. On performing optimal measurement on the meter the quantum Fisher information obtained from the meter converges to the total quantum Fisher information of an unbiased estimation considering all statistics,  {i.e.,} $F_{m}/F_{T} \to 1$ up to second order corrections in the coupling parameter $g$ (see Appendix C for the detailed derivation) \cite{Alves}. The inequality above implies that the amplification $|A_{w}|$ is constrained to not be too large for us to operate in the weak coupling regime. This is not a practical difficulty in our proposal since $g\ll 1$, as discussed below.

\begin{figure}[t]
\begin{center}
\includegraphics[width=0.44\textwidth]{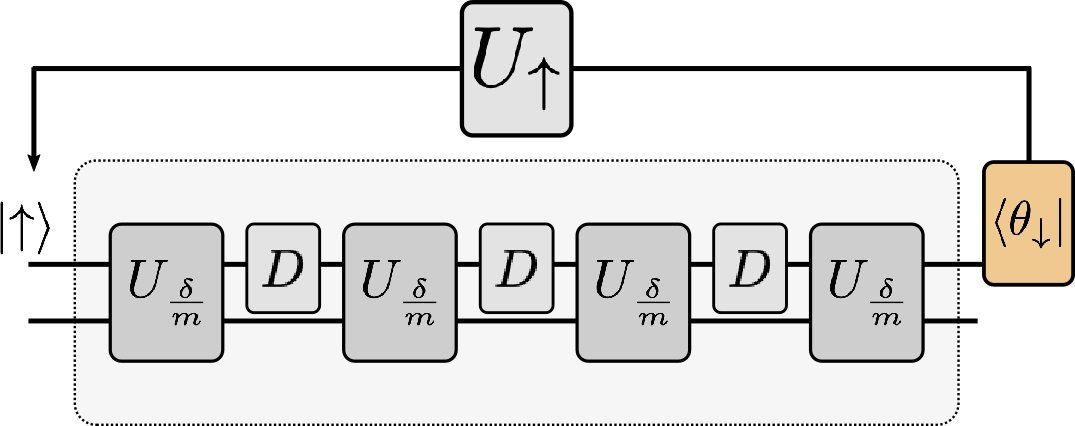}
\end{center}
\caption[Weak Magnetometer]{\emph{Decoherence, WVA \& Magnetometry:}
In the above figure, we depict concatanating the interaction unitary $U_{\frac{\delta}{m}}$ with $m$ dynamical decoupling sequences $D$, which aid in the removal of noise from the acquired signal. Following the post-selection in the state $\ket{\theta_{\downarrow}}$, we rotate the quantum state back to $\ket{\uparrow}$ (say) using fast carrier wave transitions that allow us to continue collecting signal from the qubit. Such multiple post-measurement kicks can accumulate on the meter state, depicted as the lower line in the figure and can be subsequently measured.}
\label{fig-nots}
\end{figure}

\emph{WVA for trapped ions.---} {Following recent realizations of the WVA using atomic
 systems \cite{shomroni2013demonstration,Pan2019,Wu2018},  {we consider a $^{40}$Ca$^{+}$ ion trapped in a linear Paul}
 trap as our bipartite system. The Hamiltonian of the internal qubit transition  {(taken to the two levels of the $S_{1/2}(m_{J}=\pm 1/2)$) as a qubit)}. The vibrational states of the ion are governed by the first term in the Hamiltonian (Appendix A provides a summary)
\begin{equation}
    H = \hbar\omega_{t}(a^{\dagger}a + 1/2) + \frac{\hbar \omega_{e}}{2}\sigma_{z}+\frac{\hbar g}{2c}\sigma_{z},
\end{equation}
where $\omega_{t}$ is the corrsponding frequency of the trap and $\omega_{e}$ is the energy difference of the two qubit levels. Note that we do not consider the effects of the Earth rotation, but focus only on the spin-gravity coupling. We introduce the corresponding
frequency $\omega_g\defeq g/c\approx 10^{-8}$Hz. The internal qubit degrees of freedom are coupled to the vibrational modes
 through a laser interaction, which after the usual rotating wave approximation gives the  standard Jaynes-Cummings interaction Hamiltonian
\begin{equation}
    V_{I} = \hbar \lambda (e^{-i\delta}\sigma_{+}a + e^{i\delta}\sigma_{-}a^{\dagger}),
\end{equation}
where $\sigma_\pm$ are the spin raising and lowering operators and $a$ is the annihilation operator for the vibrational mode. Furthermore $ \lambda = \eta\Omega/2 $, where $\eta$ is the Lamb-Dicke parameter, $\Omega$ the Rabi frequency of the interacting laser.
 We consider a small detuning such that $\delta = \omega_{g}t \ll 1$ which allows us to expand the exponential to the first order as
\begin{equation}
    V_{I} \approx \hbar \lambda [(1 - i\delta)\sigma_{+}a + (1 + i\delta)\sigma_{-}a^{\dagger}].
\end{equation}
We consider typical a value of $\lambda = 0.5$ kHz as the parameter for our calculation \cite{shomroni2013demonstration,Wu2018,ivanov2016high}.

 This Hamiltonian $V_{I}$ generates time evolution which can be approximated as
\begin{equation}\label{small_eq}
    U(t) \approx \prod_{m=1}^{11}U_m(t),
\end{equation}
where $U_m(t)=\exp(C_m/m!)$ is the $m^{th}$ order product in the Zassenhaus decomposition \cite{suzuki1977convergence,Casas,kimura2017explicit} and the operators $C_m$ are given in the Appendix B. Unlike the typical application of Zassenhaus formula, one of
the two terms in $V_I$ is much larger than the other term (we choose $t^*\approx6.28$ ms such that $\lambda t^*=\pi$), so we take enough terms to make sure that there is convergence from the factorial in the denominator of the Zassenhaus approximation (Appendix B). This produces an effective unitary given by
\begin{equation}
U(t^*)\approx e^{-\frac{z}{2}\lambda t^*\omega_g t^*(\sigma_{+}a-\sigma_{-}a^{\dagger})},
\end{equation}
where $z\approx-4.448$. The system is initially prepared in the \tcblack{product} state of the qubit in the excited state and the vibrational mode in the motional ground state,
 namely $\ket{\Psi_{i}}=\ket {\uparrow}_{S}\otimes \ket {0}_{M}$. On \tcblack{the} post-selecting with
  $ \ket {\theta_\downarrow}=\cos(\theta)\ket{\downarrow}_S+\sin(\theta)\ket{\uparrow}_S$ at $t^*$, we get the effective unitary acting on the vibrational state,
  \tcblack{resulting in the coherent state} \begin{equation}\label{kicked_state}
    \ket {\phi_f} = e^{\frac{z}{2}t^{*}\lambda\delta^{*} A_{w}a^{\dagger}}\ket{0}_{M},
\end{equation}
 \tcblack{where} $A_w=\bra{\theta_\downarrow}\sigma_-\ket{\uparrow}_S/\bra{\theta_\downarrow}\uparrow\rangle_S$. \tcblack{The weak value amplification that is} associated with $\delta^{*} = \omega_{g}t^{*}$
can be interpreted as a displacement operator on the vacuum-state $\ket{0}_{M}$ generating a coherent state $\vert \phi_f\rangle$. We note that since the amplitude of the coherent
  state is proportional to the unknown scale $k$, our method represents a  {broadband} magnetometer that can detect unknown small magnetic fields over several orders of magnitude by simply tuning the weak value strength.

\emph{Decoherence \& Flywheeling.---} The ideal WVA scheme works on the premise that there is no decoherence in either system or the meter.  Motional heating, laser intensity fluctuations and magnetic field noise are typical sources of decoherence
for trapped ions \cite{Decoherence2,Decoherence1}. Motional heating is not significant in our proposed \tcblack{set-up, since} on average it produces one phonon per 100 ms.
Furthermore, in a cryogenic setting reheating adds one phonon per 500 ms to the vibrational mode \cite{Schmidt_Kaler_2003,Cryogenic_ion_trap}.

  A  $1/f$ noise \tcblack{that is}
 present over the a band around the target frequency $\omega_g$ models a stray magnetic field noise. This noise  produces decoherence \tcblack{that} \tcblack{could}
 degrades the qubit signal, further deteriorating the quality of the post-measurement meter state proposed in Eq.~(\ref{kicked_state}).
  This \tcblack{decoherence} can be \tcblack{reduced by} a concatenation of dynamical decoupling sequences applied to the qubit using carrier wave transitions combined with WVA kicks at the appropriate time.

  As a simple demonstration of this strategy, we consider a Jaynes-Cummings qubit in a thermal bath and apply dynamical decoupling schemes \cite{Viola,Uhrig}.
   {Though our scheme is not optimised for the Jaynes-Cummings model, we see fidelity of 1 between the target time-evolved state (without decoherence and dynamical decoupling) and the real time-evolved state (with decoherence and dynamical decoupling applied to it) at $t^*$.}

     The Husimi-Kano Q-representation function \cite{HusimiQ,QMPhaseSpace} of the kicked vibrational state at $t^*$ is presented in Fig.~(\ref{fig-husimi}). \tcblack{It is also compared with the protocol where the WVA} is performed at  $t^*$ with no dynamical decoupling.
    While the weak value is $A_w=10^8$ for the case where the amplifying measurement is performed in presence of the dynamical decoupling sequence,
    the weak value in the absence of the dynamical decoupling sequence is \tcblack{only} $A_w=3.12$. {This demonstrates the need for our hybrid strategy combining dynamical decoupling and weak value amplification techniques.}

    Furthermore,  Fig.~\ref{fig-husimi} also \tcblack{shows} the effect of two consecutive kicks \tcblack{allow to} accumulate the effect of the weak value on the vibrational mode.
      \tcblack{Even if the dynamical decoupling sequence has not been optimised to the state of the vibrational modes there is an accumulation of the signal}.
      This demonstrates that and it is possible to detect a very small magnetic field even in the presence of a decoherence model acting on the qubit state.
       Optimizing over the bath spectral density and the Jaynes-Cummings Hamiltonian is only expected to produce better fidelities. Other noise sources such as laser field fluctuations
       are typically smaller than the field fluctuation terms and can be suppressed with similar dynamical decoupling sequences \cite{Puebla_2016}.

\begin{figure}[t]
\begin{center}
\includegraphics[width=0.5\textwidth]{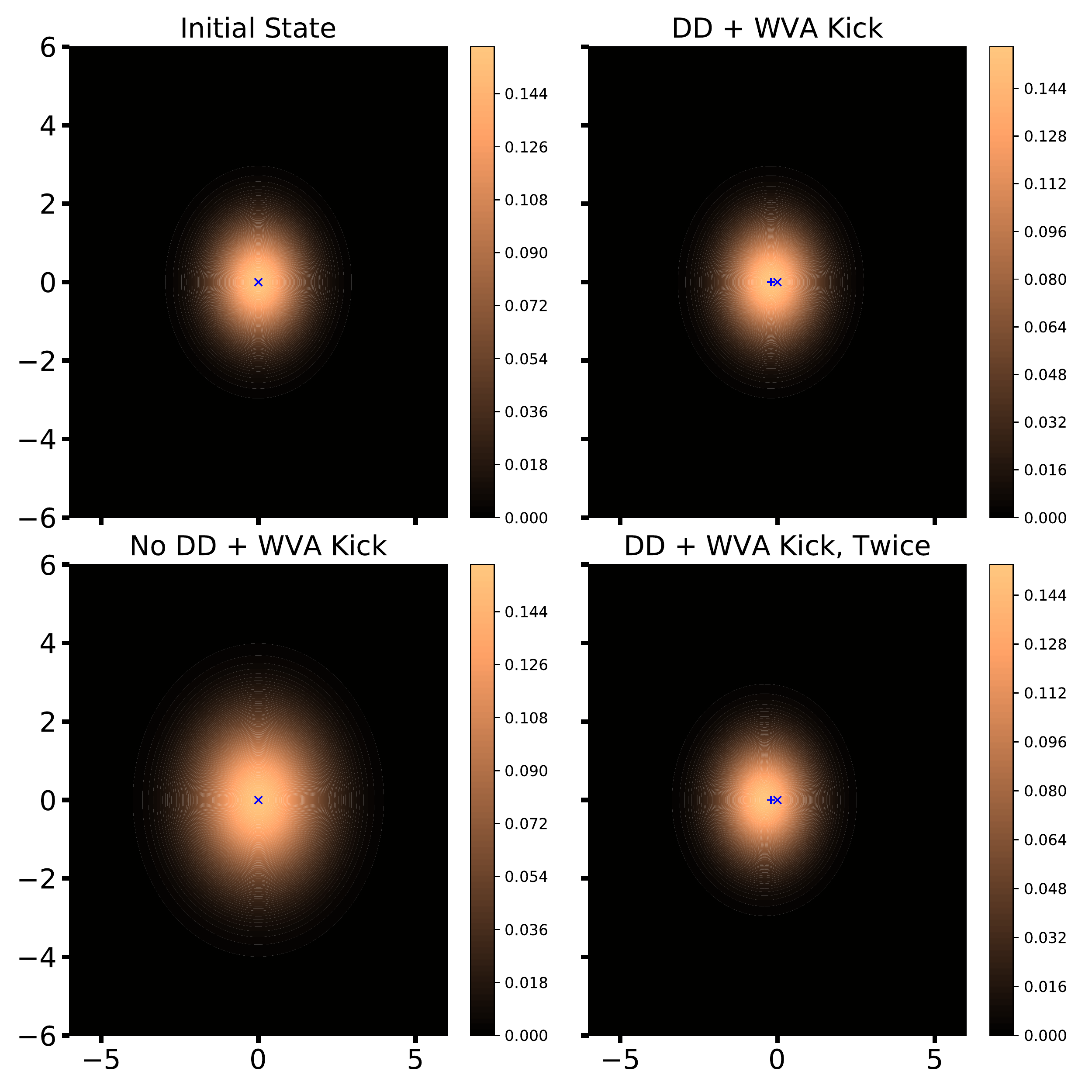}
\end{center}
\caption[Kicks]{\emph{Vibrational Husimi-Q functions:} The vibrational Husimi-Q functions are presented for (a) the initial ground state of the vibrational mode, (b) the vibrational state after one kick following a full dynamical decoupling sequence as presented in the text evolving under an extremely weak magnetic field($\times$ marks the centre of the initial state), (c) the vibrational state after two kicks evolving under an extremely weak magnetic field ($+$ marks the centre of the state with one kick) and (d) the vibrational state after one weak value kick without the dynamical decoupling sequence. Note that the vibrational state without decoupling gives rise to an amplification of $3.12$, and hence practically does not give us any information about the magnetic field.}
\label{fig-husimi}
\end{figure}

Since we have a bipartite system, we can not only induce the weak-value amplification of the signal onto the meter state, but use the meter as a flywheel \cite{ldk:16,flywheel_Poschinger} to accumulate repeated kicks.
 We can perform a flywheeling effect on the given system if through carrier transitions we return the internal states to $\ket {\uparrow}_S$.
  The dynamical decoupling sequence needs to be engineered to incorporate the effects of the vibrational state for the second kick, which is no longer in the vacuum ground state, $\ket{0}_{M}$.
  By repeating the previous steps we can obtain additional WWA effects upon the system. If we repeat the process $N$ times with the optimized dynamical decoupling sequences, we  obtain the motional state as
\begin{equation}
    \ket {N\phi_{f}} = e^{ \frac{Nz}{2}t^{*}\lambda\delta A_{w}a^{\dagger}}\ket{0}_{M}.
\end{equation}

\emph{Conclusions.---}
The WVA provides a method to use a trapped ion to detect an exceedingly small magnetic field under realistic noise assumptions. We apply this result
to detect signatures of physics beyond the Standard Model. For the SME spin-gravity coupling  corresponding  frequency is expected to be of the order of approximately $10^{-7}$Hz. Besides the specific application for searching  the terms postulated by SMEs, this method to detect extremely weak magnetic fields represents a practical technique in future quantum metrology. By employing dynamical decoupling alongside weak value amplification, we have demonstrated that the increase in the sensitivity of a practical
     detector can be enhanced in the presence of realistic noise models.
      
      Extending our analysis to include $1/f$ noise and other models of decoherence will lead to newer more sensitive practical quantum metrology techniques. 
      Besides magnetometers, techniques can be readily adapted to enhance the sensitivity of accelerometers and gyroscopes heralding a whole new range of precision measurements.

%%--------------------------------------------------------------------
%%--------------------------------------------------------------------

\emph{Acknowledgements.---}
 KLC is supported by the National Research Foundation and Ministry of Education Singapore. The work of DRT is supported by the grant FA2386-17-1-4015 of AOARD of the US Air Force. SV acknowledges support from a DST-SERB Early Career Research Award (ECR/2018/000957). Part of this work was done by SV in Nordita during the program on ``new directions in quantum information". We thank Marcus Aspelmeyer, Saikat Ghosh, Yaron Kedem, T. S. Mahesh, Robert Mann, Manas Mukherjee, Klaus M{\o}lmer, Amos Ori, Umakant Rapol and Kilian Singer for useful insights and discussions.
%---------------------------------------------------------------------------------------------------------------------------------------
%\bibliography{reference,Dani_ref,ref}

\providecommand{\noopsort}[1]{}\providecommand{\singleletter}[1]{#1}%\providecommand{\noopsort}[1]{}\providecommand{\singleletter}[1]{#1}%\providecommand{\noopsort}[1]{}\providecommand{\singleletter}[1]{#1}%
%

%---------------------------------------------------------------------------------------------------------------------------------------
\newpage
\onecolumngrid
\appendix

\section*{A:~Hamiltonian}
The total Hamiltonian of an ion trapped in a linear Paul trap can be written \cite{leibfried2003} as
\begin{equation}
    H = H_{e} + H_{m} + H_{I},
\end{equation}
where $H_{e}$ is the internal (qubit) Hamiltonian which can be expressed as ${\hbar(\omega_{e}+\omega_{g})}\sigma_{z}/{2}$, with $\hbar\omega_{e}$ is the energy difference between the qubit levels and $\omega_{g}=g/c$. $H_{m}=\hbar\omega_{t}(a^{\dagger}a + 1/2)$ is the motional
  Hamiltonian in one of the trap axis with $\omega_{t}$ the corresponding frequency of the trap potential.
   $V_{I}$ is the induced interaction between the motional and the internal states by the applied laser light,
\begin{equation}
    H_{I} = \frac{\hbar}{2}\Omega(\sigma_{+} + \sigma_{-})\Big(e^{i(kx - \omega_{l}t + \phi)} + e^{-i(kx - \omega_{l}t + \phi)}\Big).
\end{equation}
Here $\Omega$ is the Rabi frequency and $\omega_{l}=|k|c$ is the frequency of the applied laser light.
Shifting to the the interaction picture via the transformation $U(t)=\exp\big[({-i}/{\hbar})(H_{e} + H_{m})t\big]$ and performing the standard rotating wave approximation (RWA), we obtain
\begin{equation}
    V_{I}\equiv U(t)^{\dagger}H_{I}U(t) = \frac{\hbar}{2}\Omega\big[\sigma_{+}e^{i\eta(ae^{-i\omega_{t}t}+a^{\dagger}e^{i\omega_{t}t})}e^{i(\phi-\nu)} + \mathrm{h.c.} \big].
\end{equation}
Here we have already assumed to be in the Lamb-Dicke regime, $\eta^{2}(2n+1) \ll 1$ where $\eta$ is the Lamb-Dicke parameter, and $\nu = \omega_{e}t + \omega_gt - \omega_{l}t $ is the
 detuning. Performing another RWA and setting the detuning to $\nu = -\omega_{t} +\omega_g$ gives the  standard Jaynes-Cummings interaction Hamiltonian,
\begin{equation}
    V_{I} = \hbar \lambda (e^{-i\omega_gt}\sigma_{+}a + e^{i\omega_gt}\sigma_{-}a^{\dagger}), \label{eq4}
\end{equation}
with $ \lambda = \eta\Omega/2 $.

\section*{B:~Zassenhaus Expansion of Time Evolution Unitary}

Expanding the exponential of the Hamiltonian in Eq.~\eqref{eq4}
\begin{equation}
V_{I} \approx \hbar \lambda [(1 - i\omega_{g}t)\sigma_{+}a + (1 + i\omega_{g}t)\sigma_{-}a^{\dagger}].
\end{equation}
This generates the unitary operator $U(t^*) = e^{-\frac{i}{\hbar}\int_{0}^{t^{*}}dt H}$, or
\begin{equation}
U(t^*) = e^{-i\lambda t^{*}(\sigma_{+}a + \sigma_{-}a^{\dagger}) -\frac{\lambda t^{*}\omega_{g}t^{*}}{2}(\sigma_{+}a - \sigma_{-}a^{\dagger})}.
\end{equation}
This can be simplified as $U(t^*)= e^{ X+Y}$, where $X = -i\lambda t^*\big(\sigma_{+}a + \sigma_{-}a^{\dagger}\big)$ and $Y=\big(-i\lambda t^*\big)\big(\omega_{g}t^{*}\big)(\sigma_{+}a - \sigma_{-}a^{\dagger})/2$.

The Zassenhaus formula given as \cite{suzuki1977convergence,kimura2017explicit}
\begin{equation}
e^{(A+B)}=e^{A}e^{B}e^{-\frac{[A,B]}{2!}}e^{\frac{[A,[A,B]]}{3!}}e^{\frac{-[[[A,B],A],A]}{4!}} \ldots,
\end{equation}
Due to the factor $\omega_{g}t^{*}\ll 1$ in $Y$ the terms with more than one factor $Y$ in commutators can be ignored in the expansion of $U(t^*)$.
 We note that $e^{X}=e^{-i\lambda t(\sigma_{+}a + \sigma_{-}a^{\dagger})}$ is the Rabi flopping generating unitary.  At the time $t^{*} = \pi/\lambda$ and its integer multiples this operator is the identity. Hence in our calculations we consider post selecting at $t=t^{*}$.

We now present the terms in the Zassenhaus expansion.

\paragraph{Quadratic terms in $X$:}
To determine the second order correction $e^{-[X,Y]/2}$ we calculate
\begin{equation}
e^{-[X,Y]/2}=e^{\frac{i(\lambda t^{*})^{2}\omega_{g}t^{*}}{2!2}\big(\sigma_{z}(2\hat{n} + 1_{M}) + 1_{S}1_{M}\big)}.
\end{equation}
\paragraph{Cubic terms in $X$:}
The cubic term is
\begin{equation}
e^{\frac{-(\lambda t^{*})^{3}\omega_{g}t^{*}}{3!2}\big(-3\sigma_{+}a -4\sigma_{+}\hat{n}a  + 3\sigma_{-}a^{\dagger} +4\sigma_{-}a^{\dagger}\hat{n}\big)}
\end{equation}
\paragraph{Quartic term in $X$:}
We evaluate the only non-trivial the fourth order correction $e^{\frac{-[[[X,Y],X],X]}{4!}}$ as
\begin{equation}
e^{\frac{-[[[X,Y],X],X]}{4!}}=e^{\frac{i(\lambda t^{*})^{4}\omega_{g}t^{*}}{4!2}\big(3\sigma_{z}(2\hat{n} + 1_{M}) + 3(1_{S}1_{M})+8\sigma_{-}\sigma_{+}a^{\dagger}\hat{n}a +4\sigma_{+}\sigma_{-}(2\hat{n}+2\hat{n}^{2})\big)}.
\end{equation}
We note that the even order terms of $X$ are always products of equal powers of $\sigma_{-}$,$\sigma_{+}$ and $a$ and $a^{\dagger}$. Both the spin component of the system ket and the Fock state of the vibrational modes will be either eigenkets of these types of product operators or will return zero. Furthermore, the Zassenhaus expansion to any order of commutators can be written to first order in $\omega_g t$ as
\begin{equation}
U(t)=\prod_{k=1}^{K}e^{-i\frac{(\lambda t^*)^{k}(\omega_g t)}{k!}Z_k}\approx1-i\sum_{k=1}^{K}\frac{(\lambda t^*)^{k}(\omega_g t)}{k!}Z_k.
\end{equation}
We note that in the main text, we defined $C_k:=-i(\lambda t^*)^{k}(\omega_g t)Z_k$. Terms where either $Z_k\ket{\Psi}_i=0$ or $Z_k\ket{\Psi}_i=\ket{\Psi}_i$ can be readily omitted as they at most contribute to the global phase. We state the next four odd expansions of $X$ which are the only non-trivial terms that contribute to above. \\
i.)  The 5th order term in $X$ being
\begin{equation}
e^{\frac{-(\lambda t^{*})^{5}\omega_{g}t^{*}}{5!2}\big(3\{-\sigma_{+}(2\hat{n} + 1_{M})a - \sigma_{+}a(2\hat{n} + 1_{M}) +\sigma_{-}(2\hat{n} + 1_{M})a^{\dagger} + \sigma_{-}a^{\dagger}(2\hat{n} + 1_{M})\} + 2^{4}\{\sigma_{+}(\hat{n} + 1_{M})\hat{n}a-\sigma_{-}a^{\dagger}(\hat{n} + 1_{M})\hat{n} \} \big)}.
\end{equation}
ii.)  The 7th order term in $X$ being
\begin{equation}
e^{\frac{(\lambda t^{*})^{7}\omega_{g}t^{*}}{7!2}\big(12\{-\sigma_{+}a(2\hat{n} + 1_{M})a - \sigma_{+}(1_{M}+\hat{n})(2\hat{n} + 1_{M})a +\sigma_{-}(2\hat{n} + 1_{M})\hat{n}a^{\dagger} + \sigma_{-}a^{\dagger}(2\hat{n} + 1_{M})(\hat{n} + 1_{M})\} + 2^{6}\{\sigma_{+}(\hat{n} + 1_{M})^{2}\hat{n}a-\sigma_{-}a^{\dagger}(\hat{n} + 1_{M})^{2}\hat{n} \} \big)}.
\end{equation}

iii.)  The 9th order term in $X$ being
\begin{equation}
e^{\frac{-(\lambda t^{*})^{9}\omega_{g}t^{*}}{9!2}\big(48\{-\sigma_{+}a(2\hat{n}^{2} + 1_{M})a - \sigma_{+}(1_{M}+\hat{n})^{2}(2\hat{n} + 1_{M})a +\sigma_{-}(2\hat{n} + 1_{M})\hat{n}^{2}a^{\dagger} + \sigma_{-}a^{\dagger}(2\hat{n} + 1_{M})(\hat{n} + 1_{M})^{2}\} + 2^{8}\{\sigma_{+}(\hat{n} + 1_{M})^{3}\hat{n}a-\sigma_{-}a^{\dagger}(\hat{n} + 1_{M})^{3}\hat{n} \} \big)}.
\end{equation}
iv.)  The 11th order term in $X$ being
\begin{equation}
e^{\frac{-(\lambda t^{*})^{11}\omega_{g}t^{*}}{11!2}\big(192\{-\sigma_{+}a(2\hat{n}^{3} + 1_{M})a - \sigma_{+}(1_{M}+\hat{n})^{3}(2\hat{n} + 1_{M})a +\sigma_{-}(2\hat{n} + 1_{M})\hat{n}^{3}a^{\dagger} + \sigma_{-}a^{\dagger}(2\hat{n} + 1_{M})(\hat{n} + 1_{M})^{3}\} + 2^{10}\{\sigma_{+}(\hat{n} + 1_{M})^{4}\hat{n}a-\sigma_{-}a^{\dagger}(\hat{n} + 1_{M})^{4}\hat{n} \} \big)}.
\end{equation}
We terminate the series at the eleventh term since the next odd term has a prefactor of order $10^{-4}$ which ensures that it and follwing terms are much smaller relative to the first term.

\paragraph{Operation of the Zassenhaus terms:}
We consider the operation of the Zassenhaus expanded unitary $U(t)$ on the initial ket $\ket {\uparrow}_{S}\ket {0}_{M}$. We do not consider the even order terms of $\lambda$ since their operation induces only a global phase.
Without speciifying the $\sigma_{+}$ terms since $\sigma_{+}\ket {\uparrow}_{S}=0$, expanding the product of exponential terms to first order of $\omega_{g}$
\begin{multline}
1_{SM}+ \frac{(\lambda t^{*})\omega_{g}t^{*}}{2}\Big(\sigma_{-}a^{\dagger} - \frac{(\lambda t^{*})^{2}}{3!}3\sigma_{-}a^{\dagger}+ \frac{(\lambda t^{*})^{4}}{5!}\big(3\sigma_{-}\{(2\hat{n}+1_{M}),a^{\dagger}\}-16a^{\dagger}\hat{n}(\hat{n}+1_{M})\big)+\\ \frac{(\lambda t^{*})^{6}}{7!}\sigma_{-}\big(12(2\hat{n}+1_{M})\hat{n}a^{\dagger}+12a^{\dagger}(2\hat{n}+1_{M})(\hat{n}+1_{M})-2^{6}a^{\dagger}(\hat{n} + 1_{M})^{2}\hat{n}\big)-\\ \frac{(\lambda t^{*})^{8}}{9!}\sigma_{-}\big(48(2\hat{n}+1_{M})\hat{n}^{2}a^{\dagger}+48a^{\dagger}(2\hat{n}+1_{M})(\hat{n}+1_{M})^{2}-2^{8}a^{\dagger}(\hat{n} + 1_{M})^{3}\hat{n}\big)\\+\frac{(\lambda t^{*})^{10}}{11!}\sigma_{-}\big(192(2\hat{n}+1_{M})\hat{n}^{3}a^{\dagger}+192a^{\dagger}(2\hat{n}+1_{M})(\hat{n}+1_{M})^{3}-2^{10}a^{\dagger}(\hat{n} + 1_{M})^{4}\hat{n}\big)\Big)\ket {\uparrow}_{S}\ket {0}_{M}
\end{multline}
Which considering the operation on $\ket {\uparrow}_{S}\ket {0}_{M}$  is equivalent to
\begin{equation}
1_{SM}+ \frac{(\lambda t^{*})\omega_{g}t^{*}}{2}\Big(\sigma_{-}a^{\dagger} - \frac{(\lambda t^{*})^{2}}{3!}3\sigma_{-}a^{\dagger}+ \frac{(\lambda t^{*})^{4}}{5!}12\sigma_{-}a^{\dagger}+ \frac{(\lambda t^{*})^{6}}{7!}48\sigma_{-}a^{\dagger}- \frac{(\lambda t^{*})^{8}}{9!}192\sigma_{-}a^{\dagger}+\frac{(\lambda t^{*})^{10}}{11!}768\sigma_{-}a^{\dagger}\Big)\ket {\uparrow}_{S}\ket {0}_{M}.
\end{equation}
writing the sum of the Zassenhaus terms as $z=-4.44832$ in Eqn(17), the effective unitary at $t^{*}$ is
\begin{equation}
U(t)=1_{SM}+ \frac{z\lambda t^{*}\omega_{g}t^{*}}{2}\sigma_{-}a^{\dagger}.
\end{equation}

\section*{C:~Quantum Fisher Analysis of Post-selected Data}
Since the post-selection procedure discards a lot of the joint system-meter states, it is important to know how much of the information that was initially available to us in the evolved system-meter state.
  Following \cite{Davidovich}, we consider WVA with a given Hamiltonian $H$ and a given initial system-meter state $\ket{\Psi_i}$.
   We seek to know if the total quantum Fisher information, present in the initial state is still present in the kicked meter state.
   If a quantum state is measured with a fixed POVM and yields a probability distribution $P_j(\gamma):=\bra{\Psi(\gamma)}E_j\ket{\Psi(\gamma)}$, the classical Fisher information that is associated
    with this probability distribution is given by
\begin{equation}
    F_C(\gamma)=\sum_{j}\frac{1}{P_j(\gamma)}\Bigg[\frac{d P_j(\gamma)}{d \gamma}\Bigg]^2.
\end{equation}
Optimizing over the all positive operator-valued measures (POVMs) $\{E_j\}$, $E_j\geqslant 0$, $\sum_j E_j=1$, results in
 quantum Fisher information
\begin{equation}
 F_Q(\gamma)=4(\langle H^2\rangle-\langle H \rangle^2).
\end{equation}
At $t=t^{*}$ we can write our effective Hamiltonian is
\begin{equation}
    H=  -iz\gamma(\sigma_{+}a - \sigma_{-}a^{\dagger}) = -z\gamma H_{g},
\end{equation}
with $\gamma = \lambda t^{*}\omega_g t^{*}/2$ and $z$ being the Zassenhaus constant. We note that $\gamma \approx 10^{-11}$. We determine the total quantum Fisher information for the parameter $\gamma$ with the initial state $\ket{\Psi_{o}} =\ket{\uparrow}_{S}\ket{0}_{M}$ is $F_{T}(\gamma) = 4\bra{\Psi_{o}}\delta H^{2}_{o}\ket{\Psi_{o}} = 4z^{2}$.

Following the post-selection the quantum Fisher information $F_{ps}$  has two contributions
$F_{ps} = F_{m} + F_{p_f}$, where $F_{m}$ is the quantum Fisher information from the meter and $F_{p_f}$ is the classical Fisher information that can be derived from the post-selected probability distribution $p_{f}(\gamma) = \vert\vert \bra{\psi_{f}}U(\gamma)\ket{\Psi_{i}}\vert\vert^{2}$. The former can be obtained just from the meter state following an optimal measurement of the meter following  post-selection according to the protocol determined in \cite{Davidovich}.
\begin{equation}
    F_{m}(\gamma) = 4p_{f}(\gamma)\Bigg[\frac{d\bra{\phi_{f}(\gamma)}}{d\gamma}\frac{d\ket{\phi_{f}(\gamma)}}{d\gamma} - \left|\frac{d\bra{\phi_{f}(\gamma)}}{d\gamma}\ket{\phi_{f}(\gamma)}\right|^{2}\Bigg],
\end{equation}
where $p_{f}(\gamma)$ is the post-selection probability at $t=t^{*}$ and $\ket{\phi_{f}(\gamma)}$ is the meter state following post-selection.
 For notational convenience we write $\ket{\Psi_{o}} = \ket{i}\ket{0}_{M}$ and final state of the system as $\bra{\psi_{f}}=\bra{f}$. This gives post-selection probability as
\begin{equation}
    p_{f}(\gamma) = \braket{f|i}^{2} + z^{2}\gamma^{2}\braket{f|\sigma_{-}|i}^{2},
\end{equation}
and post-selected meter state as $\ket{\phi_{f}(\gamma)}=\braket{f|U(\gamma)|i}\ket{0}/\sqrt{p_{f}(\gamma)}$ which gives
\begin{equation}
    \ket{\phi_{f}(\gamma)}=\frac{\braket{f|1_{M}+z\gamma\sigma_{-}a^{\dagger}|i}\ket{0}}{\sqrt{\braket{f|i}^{2} + z^{2}\gamma^{2}\braket{f|\sigma_{-}|i}^{2}}}.
\end{equation}
Note here $\braket{f|i}$ and $\braket{f|\sigma_{-}|i}$ are real numbers.

Therefore $d\ket{\phi_{f}(\gamma)}/d\gamma$ is evaluated as
\begin{equation}
    \frac{d\ket{\phi_{f}(\gamma)}}{d\gamma}= z\braket{f|\sigma_{-}|i}\braket{f|i}\frac{-z\gamma\braket{f|\sigma_{-}|i}1_{M} + \braket{f|i}a^{\dagger}}{(\braket{f|i}^{2} + z^{2}\gamma^{2}\braket{f|\sigma_{-}|i}^{2})^{3/2}}\ket{0}.
\end{equation}
Correspondingly we evaluate the terms of the $F_{m}(\gamma)$ with
\begin{equation}
    \frac{d\bra{\phi_{f}(\gamma)}}{d\gamma}\frac{d\ket{\phi_{f}(\gamma)}}{d\gamma}= \frac{(z\braket{f|\sigma_{-}|i}\braket{f|i})^{2}}{(\braket{f|i}^{2} + z^{2}\gamma^{2}\braket{f|\sigma_{-}|i}^{2})^{2}},
\end{equation}
and the second term as
\begin{equation}
    \frac{d\bra{\phi_{f}(\gamma)}}{d\gamma}\ket{\phi_{f}(\gamma)}= 0.
\end{equation}
Therefore the quantum Fisher information that we obtain from the meter state alone,
\begin{equation}
    F_{m}(\gamma) = 4p_{f}(\gamma)\Bigg[\frac{d\bra{\phi_{f}(\gamma)}}{d\gamma}\frac{d\ket{\phi_{f}(\gamma)}}{d\gamma}\Bigg],
\end{equation}
which in terms of our Hamiltonian    is
\begin{equation}
    F_{m}(\gamma) = \frac{(\braket{f|i}^{2} + z^{2}\gamma^{2}\braket{f|\sigma_{-}|i}^{2})z^{2}(\braket{f|\sigma_{-}|i}\braket{f|i})^{2}}{(\braket{f|i}^{2} + z^{2}\gamma^{2}\braket{f|\sigma_{-}|i}^{2})^{2}}.
\end{equation}
It can be expressed in terms of $A_{w}=\braket{f|\sigma_{-}|i}/\braket{f|i}$ as
\begin{equation}
    F_{m}(\gamma) = \frac{z^{2}\braket{f|\sigma_{-}|i}^{2}}{(1 + z^{2}\gamma^{2}A_{w}^{2})},
\end{equation}
since our initial and final system states are $\ket{i}=\ket{\uparrow}_{S}$ and $\bra{f}=\cos{\theta}\bra{\downarrow}_{S}+\sin{\theta}\bra{\uparrow}_{S}$, respectively, with $\theta \approx 10^{-8}$ and hence $A_{w} \approx 10^{8}$.
 Since $\gamma \approx 10^{-11}$ therefore  $zA_{w}\gamma \ll 1$, and
\begin{equation}
    F_{m}(\gamma) \approx 4z^{2}\cos^{2}{\theta}(1 - \gamma^{2}A_{w}^{2}) \approx F_{T}(1 - z^{2}\gamma^{2}A_{w}^{2}).
\end{equation}
The quantum Fisher information from the post selection statistics is
\begin{equation}
	F_{p_f}(\gamma) = \frac{1}{p_{f}(\gamma)(1-p_{f}(\gamma))}\Bigg[\frac{dp_{f}(\gamma)}{d\gamma}\Bigg]^{2}.
\end{equation}
We have already calculated $p_{f}(\gamma)$ in Eqn(23). Its derivative
\begin{equation}
\frac{dp_{f}(\gamma)}{d\gamma} = 2z\gamma\braket{f|\sigma_{-}|i}^{2}.
\end{equation}
Therefore we evaluate $F_{p_f}(\gamma)$ from Eqn(32) as
\begin{equation}
	F_{p_f}(\gamma) = \frac{4z^{2}\gamma^{2}\braket{f|\sigma_{-}|i}^{4}}{(\braket{f|i}^{2} + z^{2}\gamma^{2}\braket{f|\sigma_{-}|i}^{2})(1-\braket{f|i}^{2} - z^{2}\gamma^{2}\braket{f|\sigma_{-}|i}^{2})}.
\end{equation}
Dividing numerator and denominator with $\braket{f|i}^{2}$ we get
\begin{equation}
	F_{p_f}(\gamma) = \frac{4z^{2}\gamma^{2}A_{w}^{2}\braket{f|\sigma_{-}|i}^{2}}{(1 + z^{2}\gamma^{2}A_{w}^{2})(1-\braket{f|i}^{2} - z^{2}\gamma^{2}\braket{f|\sigma_{-}|i}^{2})}.
\end{equation}
For our $\ket{f}$ and $\ket{i}$, $\braket{f|i}^{2} \approx 0$ and $z^{2}\gamma^{2}\braket{f|\sigma_{-}|i}^{2} \approx z^{2}\gamma^{2}$ and as previously noted $zA_{w}\gamma \ll 1$. Therefore we determine $F_{p_f}(\gamma)$ to be
\begin{equation}
	F_{p_f}(\gamma)\approx 4z^{2}\gamma^{2}A_{w}^{2}(1 - z^{2}\gamma^{2}A_{w}^{2})(1 - z^{2}\gamma^{2}).
\end{equation}
We see that $F_{p_f}$ is of order $\gamma^{2}$.
 We have hence shown that our procedure extracts the total quantum Fisher information available with the initial state up to second order terms in $\gamma$.

\section*{D: Decoherence \& Dynamical Decoupling}
As a proof of principle, we have modelled the decoherence in the system as a simple thermal damping on the qubit. The effect of the decoherence can be seen in the figure below, causing strong loss of fidelity between the target time-evolved state (without decoherence and dynamical decoupling) and the real time-evolved state (with decoherence and dynamical decoupling applied to it). In experiments involving trapped ions, another source of decoherence happens to be the $1/f$ dephasing noise generated by magnetic field fluctuations and laser intensity fluctuations, which can be modelled by suitable master equations and error mitigation methods well adapted to these master equations also exist \cite{PhysRevA.75.012308}. These require complex pulse sequences which need to be optimally designed according to the experimental setting. Recent simulations utilised concatenated pulse sequence to demonstrate the quantum Rabi model\cite{Puebla_2016}. Other techniques in eliminating dephasing noise at sub-hertz levels include using decoherence free subspace\cite{Lidar98} and should be considered in the actual experimental design.\\

To mitigate the effects of decoherence we perform periodic dynamical decoupling (PDD) by applying periodic $\pi_{Z}$ pulses upon the qubit. We consider equally spaced instantaneous pulses applied over $1.1125t^{*}$ interval. In Fig:\ref{fig-fidelity} we compare the fidelity of state evolving without decoherence to the dynamically decoupled state in the presence of decoherence and observe that for 1000 pulses we maintain a fidelity of 1, whereas for the state with decoherence where we have not applied any DD pulses the fidelity to the state without decoherence $t^{*}$ is $0.599$. Therefore we conclude that we manage to remove the effects of decoherence though our PDD implementation.
\begin{figure}[h!]
	\begin{center}
		\includegraphics[width=0.6\textwidth]{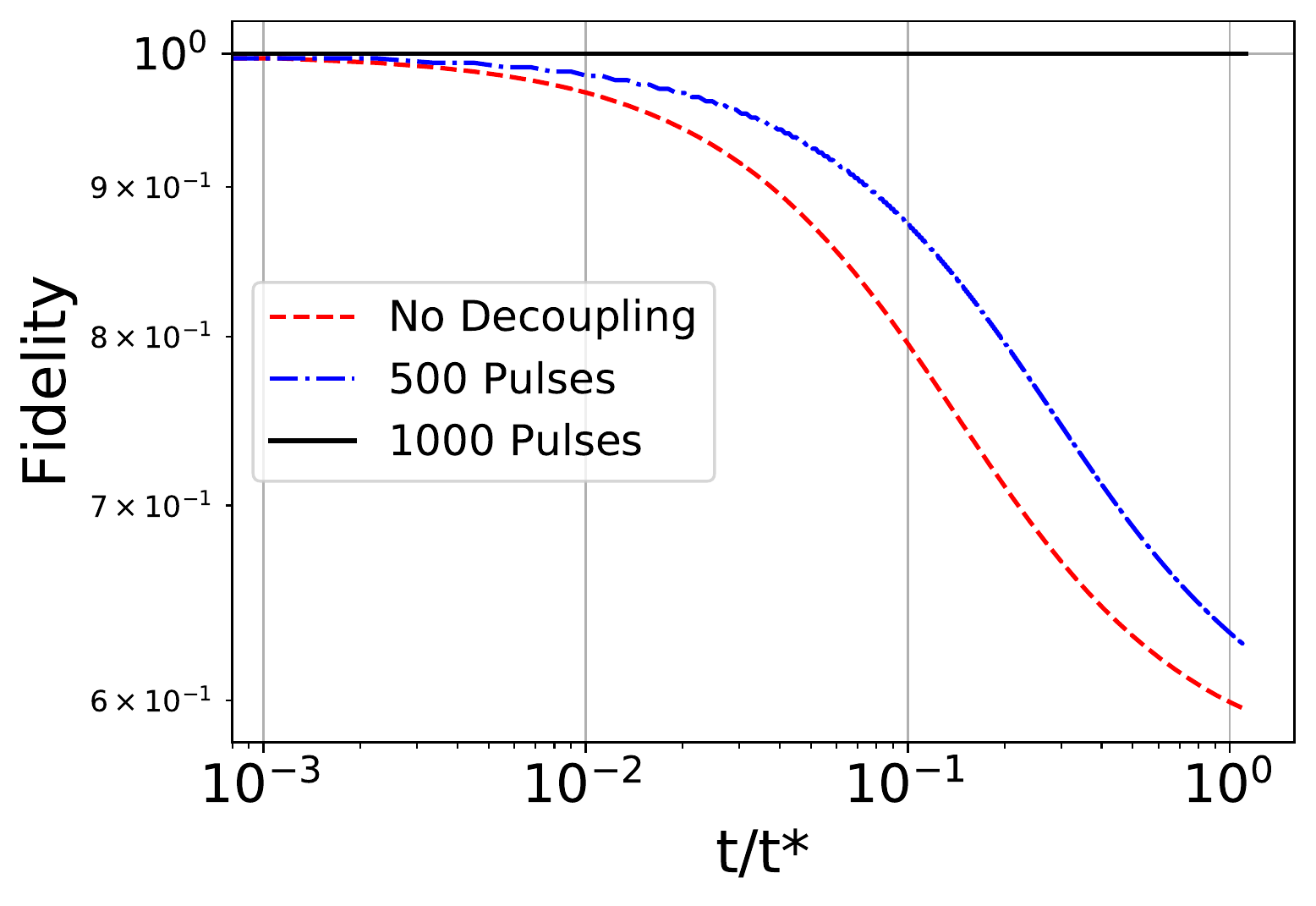}
	\end{center}
	\caption[Fidelity]{Fidelity vs. time between the target time-evolved state (without decoherence and dynamical decoupling) and the real time-evolved state (with decoherence and dynamical decoupling applied to it).}
	\label{fig-fidelity}
\end{figure}

%---------------------------------------------------------------------------------------------------------------------------------------
%\bibliography{reference,Dani_ref,ref}
\end{document}